# High-performance ferroelectric memory based on fully patterned tunnel junctions

S. Boyn[1], S. Girod[1], V. Garcia[1], S. Fusil[1], S. Xavier[2], C. Deranlot[1], H. Yamada[1,3,4], C. Carrétéro[1], E. Jacquet[1], M. Bibes[1], A. Barthélémy[1], J. Grollier[1]

[1] Unité Mixte de Physique CNRS/Thales, 1 Av. Fresnel, 91767 Palaiseau, France and Université Paris-Sud, 91405 Orsay, France

[2] Thales Research and Technology, 1 Av. Fresnel, 91767 Palaiseau, France

[3] National Institute of Advanced Industrial Science and Technology (AIST), Tsukuba, Ibaraki 305-8562, Japan

[4] JST, PRESTO, Kawaguchi, Saitama 332-0012, Japan

In tunnel junctions with ferroelectric barriers, switching the polarization direction modifies the electrostatic potential profile and the associated average tunnel barrier height. This results in strong changes of the tunnel transmission and associated resistance. The information readout in ferroelectric tunnel junctions (FTJs) is thus resistive and non-destructive, which is an advantage compared to the case of conventional ferroelectric memories (FeRAMs). Initially, endurance limitation (i.e. fatigue) was the main factor hampering the industrialization of FeRAMs. Systematic investigations of switching dynamics for various ferroelectric and electrode materials have resolved this issue, with endurance now reaching $10^{14}$ cycles. Here we investigate data retention and endurance in fully patterned submicron Co/BiFeO$_3$/Ca$_{0.96}$Ce$_{0.04}$MnO$_3$ FTJs. We report good reproducibility with high resistance contrasts and extend the maximum reported endurance of FTJs by three orders of magnitude ($4\times10^6$ cycles). Our results indicate that here fatigue is not limited by a decrease of the polarization or an increase of the leakage but rather by domain wall pinning. We propose directions to access extreme and intermediate resistance states more reliably and further strengthen the potential of FTJs for non-volatile memory applications.

Ferroelectric tunnel junctions (FTJs) consist of an ultrathin ferroelectric layer sandwiched between two different metallic electrodes[1]. Their resistance depends drastically on the orientation of the ferroelectric polarization of the tunnel barrier that is switchable with an applied electric field. This tunnel electroresistance effect can be used for non-destructive readout of the ferroelectric state.[2] FTJs were recently realized with various materials combinations.[3–10] They show potential for applications as non-volatile resistive memories with fast resistive switching[4] and large OFF/ON resistance ratios of $10^4$ (Ref. [9,10]). Moreover, owing to their memristive behavior exploiting non-uniform ferroelectric domain configurations,[6] they behave as artificial synapses and hold promise to be implemented in neuromorphic architectures.[11]

High-density integration of FTJs, as well as data retention and endurance are critical for memory applications. Up to now, submicronic FTJs were patterned as matrices of top electrodes on a continuous ferroelectric tunnel barrier/bottom electrode bilayer and were electrically connected using the conductive tip of an atomic force microscope.[4–7] Furthermore, endurance in FTJs has only been demonstrated up to a few $10^3$ cycles,[4,9,10] in contrast to the cycling performance of capacitive ferroelectric memories ($>10^{14}$ cycles).[12,13] Here, we demonstrate high-yield on series of fully-patterned submicron FTJs with long data retention in the ON and OFF states and endurance up to $4\times10^6$ cycles.

In previous studies, it has been shown that the giant change in tunnel resistance of FTJs is directly linked to the reversal of the polarization in the ferroelectric barrier.[2,4,9,10,14–16] Piezoresponse force microscopy was used to correlate the ferroelectric domain configuration in the FTJ with electronic transport measurements.[6,9,10] In the specific case of Co/BiFeO$_3$/Ca$_{0.96}$Ce$_{0.04}$MnO$_3$

(Co/BFO/CCMO) junctions,[10] the tunnel resistance is in the OFF (ON) state when the polarization of BFO points towards CCMO (Co). The OFF/ON resistance contrast is $10^4$ and mixed states of up- and down-pointing polarization result in intermediate resistance levels in which both types of domains conduct in parallel. Based on this promising system, we fabricated a chip of 50 fully patterned, macroscopically connectable FTJs, a step further towards application as memory cells.

The growth of BFO (4.6 nm)/CCMO (20 nm) on $YAlO_3$ (YAO) substrates is done by pulsed laser deposition as described elsewhere.[10,17] A Mn-doped (5 %) target is used for the growth of BFO. The fabrication process of the FTJs presented here (Fig. 1) consists of five major steps. During the first step, Pt (90 nm)/Co (5 nm) nanopillars (of 300, 400, and 500 nm in diameter) are defined by conventional e-beam lithography and lift-off on the top of the BFO/CCMO//YAO multilayer (Fig. 1c). The 100 µm wide bottom electrode is then patterned by etching the BFO/CCMO layers. In order to planarize and electrically insulate the top from the bottom electrodes of the pillar, a thin photoresist layer is deposited and polymerized. Subsequently, the top of the pillar is uncovered from the photoresist by reactive ion etching. Finally, macroscopic contact pads are defined by optical lithography, followed by Ni (20 nm)/Au (200 nm) deposition and lift-off.

The transport measurements are conducted using a Yokogawa GS610 voltage source in series with a Keithley 6487 picoammeter. This DC circuit is separated from the Agilent 81150A pulse generator by a bias tee. The FTJ contact pads (Fig. 1b) are connected using a standard RF probe, fixing ground level to the bottom electrode.

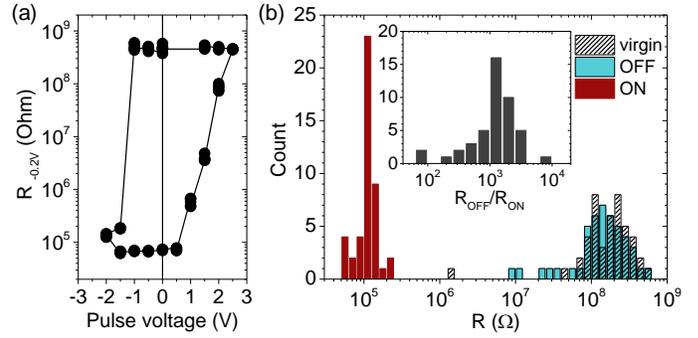

**Fig. 2. (a)** Typical resistance hysteresis of a FTJ as a function of pulse voltage. The junction is initially in the high resistive virgin state and switches to the low resistive ON state upon application of negative voltage pulses with increasing amplitude. The application of positive voltage pulses gradually increases the resistance to the OFF state which restores the resistance of the virgin state. A step of 0.5 V is chosen for the voltage pulses. **(b)** Distributions of the virgin, ON, and OFF states for the 45 working FTJs. The inset shows the distribution of the OFF/ON ratio. Resistance was measured under DC voltage of -0.2 V.

Among the 50 FTJs on the sample, 10 % are either short-circuited or unswitchable. The remaining FTJs show resistance switching between a high resistive OFF state and a low resistive ON state (see example in Fig. 2a). The OFF/ON ratio is higher than 100 for 86 % and higher than 1000 for 64 % (Fig. 2b). The virgin resistance of these FTJs is high with a mean value of $2\times10^8$ Ohm measured at a DC voltage of -0.2 V. The high resistive junctions can be switched to a low resistive ON state by successive application of 100 ns voltage pulses down to -2 V. Their ON state resistance averages at $1\times10^5$ Ohm with a low standard deviation of $3\times10^4$ Ohm. Upon successive application of 100 ns voltage pulses with increasing amplitude up to 2.5 V the resistance increases to finally recover the virgin value in the OFF state of $2\times10^8$ Ohm with a standard deviation of $1.2\times10^8$ Ohm. These statistics and the consistency with our previous results[10] on the same system show the reliability of the ferroelectric switching and the good control of the FTJ fabrication process.

When used as a memory, good retention properties of FTJs are crucial. In order to evaluate those, we used the following measurement scheme for a representative junction. The FTJ being in the low resistive ON state, a 100 ns write pulse of variable amplitude is applied. After each write pulse, the resistance at $V_{DC} = -0.2$ V is recorded in increasing time intervals up to 25 h of total measurement time (Fig. 3). The ON and OFF states, i.e., when the polarization of the BFO is in a nearly saturated configuration, show very good retention. By extrapolation, we estimate that the OFF/ON resistance ratio after 10 years would still be more than two orders of magnitude (~375) on these 300 to 500 nm wide FTJs. These good retention properties have so far only been obtained on 30 µm wide $BaTiO_3$-based FTJs and our devices represent a 100 fold improvement for scalability.[9]

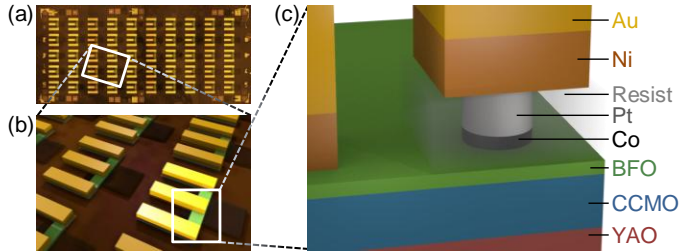

**Fig. 1. (a)** Optical microscope image of the chip after patterning showing 5×10 FTJs. **(b)** 3D representation of a zoomed area containing a few FTJs. The three parallel bars are the ground-signal-ground contact pads. **(c)** 3D sketch of one FTJ.

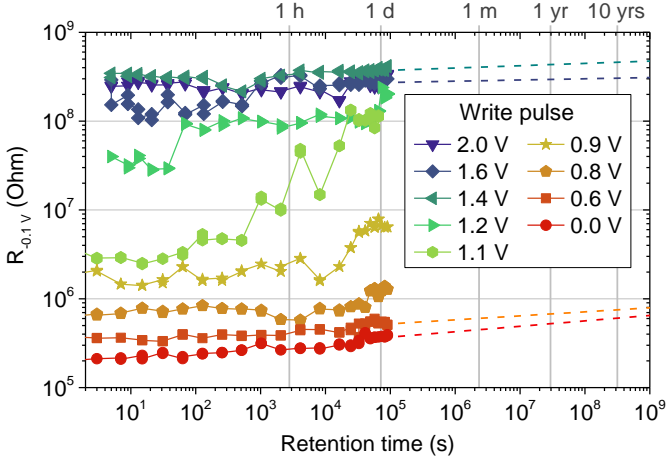

**Fig. 3.** Resistance as a function of time in different initial resistance states reached by application of write pulses of different voltages. The dashed lines show the extrapolation to $10^9$ s. Resistance was measured under DC voltage of -0.1 V.

The intermediate resistance states (write pulses of 0.8 to 1.2 V) are less stable and tend to relax towards higher resistances. The strongest relaxation is observed after application of a pulse of 1.1 V, where the resistance was initially set to $3\times10^6$ Ohm, and relaxes to $10^8$ Ohm after 25h. The domain configuration in ferroelectric thin films is governed by two energies: the contribution from the depolarizing field and the domain wall energy. On the one hand, as the thickness of the ferroelectric layer decreases, the depolarizing field increases, favoring a multi-domain polarization. On the other hand, if energy is needed for the creation of domain walls the system will tend to lower their length, which is minimized in a mono-domain state. As the nearly saturated ON and OFF states are more stable than intermediate, multi-domain states, it seems that domain wall energy is significant in our system. Additionally, the voltage threshold to destabilize the ON state is lower than the voltage threshold of the OFF state (Fig. 2a). This indicates a more energetically favorable downwards polarization state which could result from the FTJ's interfacial asymmetry.[18] The relaxation of intermediate states towards the OFF state is in agreement with this scenario.

The endurance of the FTJs was tested by applying voltages pulses of 100 ns duration. Each cycle consists of the application of 1.5 V to set the high resistance state and -1.3 V to set the low resistance state. Ideally, the resistance has to be read after each pulse, but as the picoammeter we have used can take up to 2 s to perform a measurement, this procedure implies unreasonably long measurement time (>40 days for $10^6$ cycles). We therefore measured the high and low resistance states in increasing intervals of 1, 2, 4, 8, etc. cycle trains.

We noticed that the FTJ is sometimes blocked in the high resistance state where the application of the negative voltage pulse does not lead to resistance switching. In the case of a blocked junction, we discarded the preceding train of cycles and automatically applied negative pulses of increasing amplitude (up to $V_p = 3$ V) until the resistance dropped below a defined limit of $10^6$ Ohm for the ON state resistance. After this, we proceeded by applying the above mentioned procedure of cycle trains restarting at 1 cycle.

A typical example of the endurance measurements is presented in Fig. 4. We reached up to $4\times10^6$ cycles with a resistance contrast of about two orders of magnitude. This narrowed resistance difference is due to the reduced pulse voltages which only partly switch the polarization of BFO.[10] During this experiment the resistance was blocked and unblocked 1039 times, i.e. in average every ~4000 cycles.

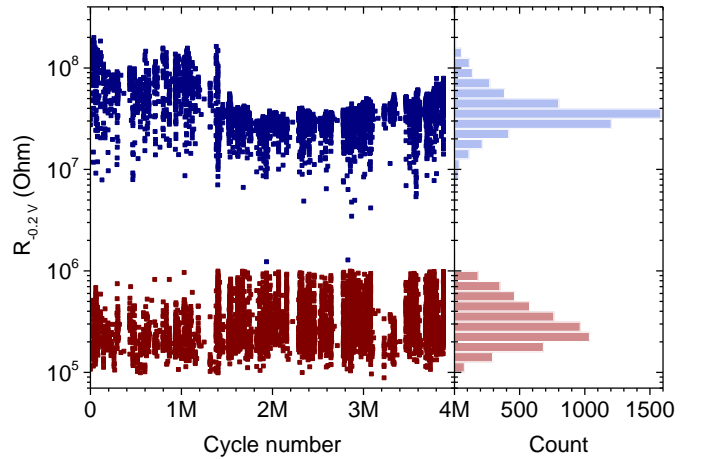

**Fig. 4.** (a) Resistance as a function of cycle number. (b) Distribution of the ON and OFF state resistances measured in (a). Resistance was measured under DC voltage of -0.2 V.

After cycling the FTJs can still be switched to the initial OFF and ON resistance states. This indicates that the switchable polarization of the BFO is not affected (in contrast to Ref. [19,20]). The switching is, however, less deterministic, indicating stronger pinning of domain walls, probably due to creation or rearrangement of defects, such as oxygen vacancies. Additionally, the low resistance state requires higher voltage pulse amplitudes than before cycling to be reached as previously observed by Baek et al.[20] This could be related to an increased internal field by the movement of oxygen vacancies[13,19,21,22] which destabilizes the ON state. A possible means of slowing down the migration of oxygen vacancies and therefore enhancing this result is doping the BFO by La or Nb to reduce their mobility.[13]

In summary, we have fabricated a chip of fully patterned FTJs with a few hundred nanometers in diameter consisting of a BFO ferroelectric tunnel barrier sandwiched between Co and CCMO electrodes. We obtained a high yield of working devices with an average OFF/ON resistance ratio of $10^3$. The ON and OFF states are highly stable, extrapolated to 10 years of data retention. Noticeably, we demonstrated endurance performance of $4\times10^6$ cycles, well above state of the art reports on FTJs. This is an important step towards FTJs integration in non-volatile memories and realization of high-density crossbar architectures.

Financial support from the European Research Council (ERC Advanced Grant No. 267579 and ERC Starting Grant No. 259068) and the French Agence Nationale de la Recherche (ANR) projects MHANN and NOMILOPS is acknowledged.

Please address correspondence to V.Garcia. Electronic mail: vincent.garcia@thalesgroup.com